\documentclass{elsart}

\usepackage[square,comma]{natbib}
\usepackage{graphicx}
\usepackage{pxfonts}
\usepackage{lineno}

\usepackage{amssymb}

\journal{}

\begin{document}

\thispagestyle{empty}
\begin{Large}
\textbf{DEUTSCHES ELEKTRONEN-SYNCHROTRON}

\textbf{\large{Ein Forschungszentrum der Helmholtz-Gemeinschaft}\\}
\end{Large}

DESY 11-244

December 2011

\begin{eqnarray}
\nonumber &&\cr \nonumber && \cr \nonumber &&\cr
\end{eqnarray}
\begin{eqnarray}
\nonumber
\end{eqnarray}
\begin{center}
\begin{Large}
\textbf{Scheme for generating and transporting THz radiation to the
X-ray experimental hall at the European XFEL}
\end{Large}
\begin{eqnarray}
\nonumber &&\cr
\end{eqnarray}

\begin{large}
Winfried Decking$^a$, Gianluca Geloni$^b$, Vitali Kocharyan$^a$,
\end{large}

\begin{large}
Evgeni Saldin$^a$ and Igor Zagorodnov$^a$
\end{large}

\textsl{\\$^a$Deutsches Elektronen-Synchrotron DESY, Hamburg}

\textsl{\\$^b$European XFEL GmbH, Hamburg}

\begin{eqnarray}
\nonumber
\end{eqnarray}
ISSN 0418-9833
\begin{eqnarray}
\nonumber
\end{eqnarray}
\begin{large}
\textbf{NOTKESTRASSE 85 - 22607 HAMBURG}
\end{large}
\end{center}
\clearpage
\newpage

\begin{frontmatter}



\title{Scheme for generating and transporting THz radiation to the X-ray experimental hall at the European XFEL}


\author[DESY]{Winfried Decking}
\author[XFEL]{Gianluca Geloni\thanksref{corr},}
\thanks[corr]{Corresponding Author. E-mail address: gianluca.geloni@xfel.eu}
\author[DESY]{Vitali Kocharyan}
\author[DESY]{Evgeni Saldin}
\author[DESY]{and Igor Zagorodnov}
\address[XFEL]{European XFEL GmbH, Hamburg, Germany}
\address[DESY]{Deutsches Elektronen-Synchrotron (DESY), Hamburg,
Germany}

\begin{abstract}
The design of a THz edge radiation source for the European XFEL is
presented. We consider generation of THz radiation from the spent
electron beam downstream of the SASE2 undulator in the electron beam
dump area. In this way, the THz output must propagate at least for
$250$ meters through the photon beam tunnel to the experimental hall
to reach the SASE2 X-ray hutches. We propose to use an open beam
waveguide such as an iris guide as transmission line. In order to
efficiently couple radiation into the iris transmission line,
generation of the THz radiation pulse can be performed directly
within the iris guide. The line transporting the THz radiation to
the SASE2 X-ray hutches introduces a path delay of about $20$ m.
Since THz pump/X-ray probe experiments should be enabled, we propose
to exploit the European XFEL baseline multi-bunch mode of operation,
with $222$ ns electron bunch separation, in order to cope with the
delay between THz and X-ray pulses.  We present start-to-end
simulations for $1$ nC bunch operation-parameters, optimized for THz
pump/X-ray probe experiments. Detailed characterization of the THz
and SASE X-ray radiation pulses is performed. Highly focused THz
beams will approach the high field limit of 1 V/atomic size.
\end{abstract}

%
%
%
\end{frontmatter}



\section{\label{sec:uno}  Introduction}

The exploitation of a high power coherent THz source as a part of
the LCLS-II user facility has been proposed in the LCLS-II CDR
\cite{CDRL2}. THz radiation pulses can be generated by the spent
electron beam downstream of the X-ray undulator. In this way,
intrinsic synchronization with the X-ray pulses can be achieved. A
first, natural application of this kind of photon beams is for the
pump-probe experiments. Through the combination of THz pump and
X-ray probe, XFELs would offer unique opportunities for studies of
ultrafast surface chemistry  and catalysis \cite{CDRL2}. Also, the
LCLS team started an $R\&D$ project on THz radiation production from
the spent electron beam downstream of the LCLS baseline undulator
\cite{GALA}-\cite{FISH}. In that case, THz pulses are generated by
inserting a thin Beryllium foil into the electron beam. In this
paper we describe a scheme for integrating such kind of radiation
source at the European XFEL facility \cite{EXFEL}.

We begin our considerations with the generation of THz radiation
from the spent electron beam downstream of the SASE2 undulator in
the electron beam-dump area. We then move to consider the transport
of THz radiation pulses from the XFEL beam dump-area to the
experimental hall. This constitutes a challenge, because the THz
output must propagate at least $250$ meters in the photon beam
tunnel and in the experimental hall to reach the SASE2 X-ray
hutches. Since THz beams are prone to significant diffraction, a
suitable beam transport system must be provided to guide the beam
along large distances maintaining it, at the same time, within a
reasonable size. Moreover, the THz beamline should be designed to
obtain a large transmission efficiency for the radiation over a wide
wavelength range. Transmission of the THz beam can only be
accomplished with quasi-optical techniques. In this paper, similarly
as in \cite{OURT}, which focused on the LCLS baseline, we propose to
use an open beam waveguide such as an iris guide, that is made of
periodically spaced metallic screens with holes, for transporting
the THz beam at the European XFEL. The eigenmodes of the iris guide
have been calculated numerically for the first time by Fox and Li
\cite{FOX1} and later obtained analytically by Vainstein
\cite{VAI1,VAI2}. In \cite{OURT} we already presented a complete
iris guide theory. In particular, the requirements on the accuracy
of the iris alignment were studied. In order to efficiently couple
radiation into the transmission line, it is desirable to match the
spatial pattern of the source radiation to the mode of the
transmission line. To this end, it is advisable to generate
radiation from the spent electron beam directly in the iris line
with the same parameters used in transmission line. In this way, the
source generates THz radiation pulses with a transverse mode that
automatically matches the mode of the transmission line. The theory
described in \cite{OURT} supports this choice of THz source.

In the present work we present a conceptual design for a THz edge
radiation source at the European XFEL. It includes an $80$ m-long
electron beam vacuum chamber equipped with an iris line, and a $250$
m-long transmission line with the same parameters. The transmission
line, which develops through the XTD7 distribution tunnel and field
$5$ of the experimental hall, includes at least ten $90$-degrees
turns with plane mirrors at $45$ degrees as functional components.
It is possible to match incident and outgoing radiation without
extra losses in these irregularities. The transport line introduces
a path delay of about $20$ m between THz and X-ray pulses generated
from the same electron bunch. Since THz pump/X-ray probe experiments
should be enabled, in order to cope with this delay we propose to
exploit the unique bunch structure foreseen as baseline mode of
operation at the European XFEL, with $222$ ns electron bunch
separation, together with an additional delay line in experimental
hall\footnote{The temporal resolution of pump-probe experiments
using THz and X-ray pulses should be limited by the duration of the
THz pulse (about $300$ fs) rather than by the jitter between the
electron bunches (about $50$ fs).}.

\section{\label{due} Principles of THz radiation generation}

The availability of any THz source at the European XFEL facility
should be complemented with the availability of a suitable THz beam
transport system, which must guide the beam for distances in the
$300$ meters range.

The THz beam can only be transmitted with quasi-optical techniques.
In particular, the idea of providing a periodic phase correction for
the free- space beam, in order to compensate for its divergence, is
very natural. In the 1960s numerous attempts of designing various
quasi-optical transmission lines were reported. In particular, it
was proposed to use open beam waveguides such as lens guides, mirror
guides, and iris guides \cite{FOX1,VAI1},\cite{GOUB}-\cite{FERN}.
The competition among different proposals ended with the victory of
mirror guides, still in use today for example for plasma heating
\cite{SORO,FERN}.

\begin{figure}[tb]
\includegraphics[width=1.0\textwidth]{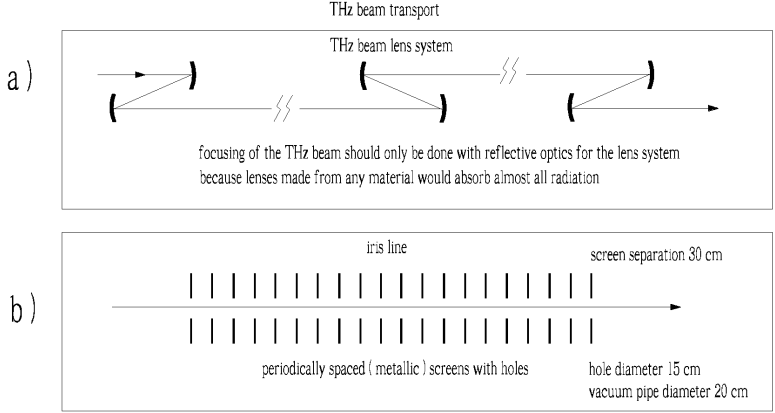}
\caption{Optical arrangement of the transport system based on the
use of (a) a mirror guide and (b) an iris guide} \label{T1}
\end{figure}
Focusing of the THz beam can only be provided with reflective optics
since lenses made of any material would reflect and absorb all
radiation at long distances. Fig. \ref{T1} (a) shows the optical
arrangement of the transport system based on the use of a mirror
guide. Each focusing unit is composed by two matched copper mirrors.
The mirrors are separated in such a way that the incident angle is
sufficiently small to minimize astigmatism. Existing mirror guides
are characterized by a maximal length of about 40-60 m
\cite{KULI,TIED}. As discussed above, at European XFEL facility
scientists need a beam transport system working for significantly
longer distances, and the advantages of a mirror-guide based
transport system are not evident.

In this paper we propose, as an alternative to the mirror line
solution, to use an iris line, Fig. \ref{T1} (b).  Iris lines are
characterized by a low attenuation of the fundamental mode,
self-filtering of high order modes, wide operational wavelength
range and mechanical integrity of the structure. The first
investigation of the influence of diffraction effects on the
formation of the field eigenmodes in an iris line was carried out by
Fox and Li using physical optics techniques \cite{FOX1}.  A very
different and mathematically solid approach to the same problem was
introduced by Vainstein \cite{VAI1,VAI2}. His studies are based on
direct solution of Maxwell equations. An analysis of diffraction and
reflection from the iris edges allowed Vainstein to derive for the
first time analytic expressions for field distribution and mode
losses in iris guides. An infinite, discrete set of eigenfunctions
with corresponding complex eigenvalues can be found. This set of
modes comprises leaky modes which are similar albeit not identical
to modes of microwave waveguide with resistive walls. For instance,
one can find that iris line modes, in contrast with microwave
waveguide leaky modes, are independent of the polarization of the
radiation.

Consider an axially symmetric iris guide with inner radius $a$ and
distance between the two screens $b$, and call
$\vec{E}(\vec{r},z,\omega)$ the temporal Fourier transform of the
transverse electric field of the radiation in the guide. Since the
transverse size of the radiation is much larger than the reduced
wavelength $\lambdabar = c/\omega$, the field envelope
$\vec{\widetilde{E}} = \vec{E} \exp[-i \omega z/c]$ turns out to be
a slowly varying function of the longitudinal coordinate $z$ with
respect to the wavelength, and it obeys the paraxial Maxwell
equation. Given the symmetry of the problem we can introduce polar
coordinates $(r,\phi)$, and seek a solution for the slowly varying
field amplitude $\widetilde{E}(r,\phi,z)$ of given polarization
component in the form

\begin{eqnarray}
\widetilde{E} = u_{nj}(r)\exp[-in \phi -i \Delta k_z z]~,
\label{solu}
\end{eqnarray}

with n = 0, 1, 2, ... . In the first order of the small parameter $M
= (8 \pi N)^{-1/2}$, where $N = a^2/(\lambda b)$ is the Fresnel
number, the functions $u_{nj}$ assume the form

\begin{eqnarray}
u_{nj} = J_n(k_{nj} r) \label{unfun}
\end{eqnarray}
where

\begin{eqnarray}
k_{nj} = \frac{\nu_{nj}}{a} [1-(1+i) \beta_0 M]~, \label{knj}
\end{eqnarray}
with\footnote{$\beta_0$ turns out to be related with one of the most
famous mathematical functions, the Riemann zeta function $\zeta$. In
fact, it is given by $-\zeta(1/2)/\sqrt{\pi}$, see
\cite{VAI1,VAI2}.} $\beta_0 = 0.824$, and $\nu_{nj}$ the $j$-th root
of the $n$-th order Bessel function of the first kind (i.e.
$J_n(\nu_{nj}) = 0$). Substituting the expression for $k_{nj}$ into
the dispersion relation

\begin{eqnarray}
k_z^2 + k_{nj}^2 = \frac{\omega^2}{c^2} \label{kzzz}
\end{eqnarray}
we obtain the following expression for $\Delta k_z$:

\begin{eqnarray}
\Delta k_z b =  - 2 \nu_{nj}^2 M^2 + 4 \nu_{nj}^2 M^3 (1+i)\beta_0~
. \label{kzzz2}
\end{eqnarray}
For the eigenmode with transverse wavenumber $k_{nj}$, the fraction
of the radiation power losses per transit of one iris is given by

\begin{eqnarray}
2 \mathrm{Im}(k_z b) = 8 \nu_{nj}^2 M^3 \beta_0 ~. \label{loss}
\end{eqnarray}
The relative loss of the $j$-th mode of order $n$ after traveling
for a distance $z$ is therefore given by

\begin{eqnarray}
\left(\frac{\Delta W}{W}\right)_{nj} = 1- \exp\left(-\frac{
\nu_{nj}^2\beta_0  }{ (2\pi N)^{3/2}}\frac{z}{b}\right) = 1-
\exp\left(-\frac{ \nu_{nj}^2\beta_0 (\lambda b)^{3/2} }{
(2\pi)^{3/2} a^3}\frac{z}{b}\right)~.\label{loss}
\end{eqnarray}
Due to the exponential dependence on $\nu_{nj}^2$, only the lower
order modes tend to survive. Note that the exponent in Eq.
(\ref{loss}) depends on the distance between two irises, $b$, only
weakly as $\sqrt{b}$, while there is a much stronger dependence on
$\lambda$ and $a$.

\begin{figure}[tb]
\includegraphics[width=1.0\textwidth]{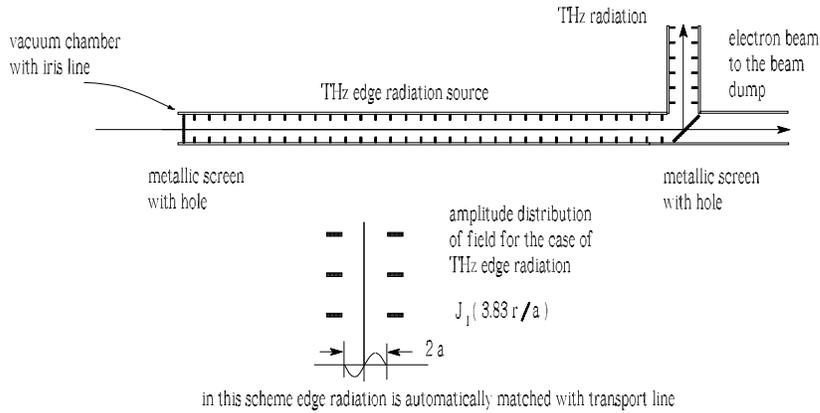}
\caption{Edge radiation is generated with the help of upstream and
downstream metallic screens.  } \label{T2}
\end{figure}
As already mentioned, an iris line constitutes a suitable THz beam
transport system for the European XFEL. For example, calculations
indicate that the losses of the principal mode at $\lambda = 100~
\mu$m, $b = 30$ cm and $2a = 15$ cm should be in the order of $10
\%$  after traveling for a distance $z = 250$ m.  The transfer line
pipe diameter is estimated to be $20$ cm.  It is therefore
technically feasible to install such transfer line inside the
distribution tunnel XTD7. However, as discussed before, in order to
efficiently couple radiation into the transmission line, one needs
to match the spatial pattern of the source radiation to the mode of
the transmission line.

Let us consider a coherent edge radiation source in the THz range,
which is relatively simple to implement at the European XFEL. A
setup where edge radiation is formed with the help of upstream and
downstream metallic screens is shown in Fig. \ref{T2}. The edge
radiation from the upstream screen is extracted by the downstream
screen, which acts as a mirror, and is sent to the iris transmission
line. The length of straight section between upstream screen and
mirror plays the role of the length of the insertion device for edge
radiation (see for example \cite{oured} for a review on these
matters). The matching problem is easily solved if the THz source is
equipped with an iris line as well \footnote{A hole may be present
in the edge radiation screens. A hole with a diameter of a few
millimeters will not perturb the X-ray beam, nor the electron beam,
nor the THz beam.}.

When dealing with a setup where THz radiation from an ultra
relativistic electron beam is extracted by a mirror, one usually
talks about Backward Transition Radiation (BTR). The main problem to
solve is the specification of the electric field distribution at
some position where the mirror is present. It should be stressed
that the specification of the field at the mirror position must be
considered as the first step to the specification of the field at
the sample position. Such first step is considered separately,
because the field at the mirror position is independent of the type
of mirror and outcoupling optics. Once the field at the mirror
position is known, the problem of specification of the field at the
sample position can be solved for example with the help of Physical
Optics techniques.

Let us discuss the problem of field characterization at the mirror
position in more detail. When electrons are in unbounded space and
come from an infinitely long straight line, the field distribution
in the mirror plane can be calculated analytically following
Ginzburg and Frank \cite{GINZ}.  In most practical cases, however,
the Ginzburg-Frank equation is not applicable because two basic
assumptions of the analytical derivation are not fulfilled: the
electron beam is moving inside the metallic vacuum chamber, which
effectively acts like an overmoded waveguide, and the straight line
has a finite length with a bending magnet at the upstream end. It is
known that Ginzburg-Frank theory is the limiting case of the more
general theory of Edge Radiation in unbounded space. For a review on
these matters see, among others, \cite{BOSC}. For a very recent
review on edge radiation and further references see, for example,
\cite{SMOL}. Emission of edge radiation in the presence of metallic
boundaries has been a much less-treated subject in literature,
compared to the unbounded space case. To the best of our knowledge,
there is only one article reporting on edge radiation from electrons
in a homogeneous metallic overmode waveguide, in particular with
circular cross-section \cite{oured}. The method described in
\cite{oured} is therefore capable of treating realistic experimental
setups. In \cite{OURT} we applied the method described in
\cite{oured} to the case of an iris guide\footnote{A very different
approach for generating coherent THz radiation from
ultra-relativistic ultrashort electron bunch in a vacuum chamber
with specially introduced roughness was proposed in \cite{NOVA}}.

To fix ideas we focus our attention on the setup in Fig. \ref{T2}.
Electrons travel through  the usual edge radiation setup. The
difference is that now we account for the presence of the iris guide
along the straight section. Since electrons pass through an upstream
edge screen, one may  assume that the iris guide starts at the
upstream screen position. One should calculate the  field
distribution at the mirror position, which should be subsequently
propagated to the experimental hall. The presence of an upstream
edge screen seems at first glance not necessary, because electrons
come in any case from the straight section. However, due to
variations of the vacuum chamber cross-section upstream of the edge
radiation setup, characterization of the field distribution at the
upstream open-end of the iris guide is problematic. In spite of
this, the electric field of the electron beam in the plane
immediately behind the upstream screen can be well defined as zero,
leading to extra simplifications. In this case we deal with a
well-defined problem, and this allow us to characterize the field
distribution at the end of the iris guide in the mirror plane.

In the case of the THz source at the European XFEL, the electron
beam transverse size is much smaller than the diffraction size. This
means that, as pertains the characterization of the THz pulses , the
electron beam can be modeled as a filament beam. In this case the
electron beam current is made up of moving electrons randomly
arriving at the entrance of the iris guide, and obeying

\begin{eqnarray}
I(t) = (-e) \sum_{k=1}^{N_\mathrm{e}} \delta(t-t_k)~,
\label{current}
\end{eqnarray}
where $\delta(\cdot)$ is the Dirac delta function, $(-e)$ is the
electron charge, $N_\mathrm{e}$ is number of electrons in a bunch,
and $t_k$ is the random arrival time of the $k$-th electron at the
undulator entrance. The electron bunch profile is described by the
profile function $F(t)$. $F(t)dt$ represents the probability of
finding an electron between time $t$ and time $t+dt$. The beam
current averaged over an ensemble of bunches can then be written in
the form:

\begin{eqnarray}
\langle I(t)\rangle = (-e) N_\mathrm{e} F(t) ~. \label{Iave}
\end{eqnarray}
The radiation power at frequency $\omega$, averaged over an
ensemble, is given by the expression:

\begin{eqnarray}
\langle P(\omega) \rangle = p(\omega) [N_\mathrm{e} +
N_\mathrm{e}(N_\mathrm{e}-1)|\bar{F}(\omega)|^2]~, \label{pow}
\end{eqnarray}
where $p(\omega)$ is the radiation power from one electron and
$\bar{F}(\omega)$ is the Fourier transform of the bunch profile
function. For wavelengths shorter than the bunch length the form
factor reduces to zero. For wavelengths longer than the bunch length
it approaches unity.

\begin{figure}[tb]
\includegraphics[width=0.5\textwidth]{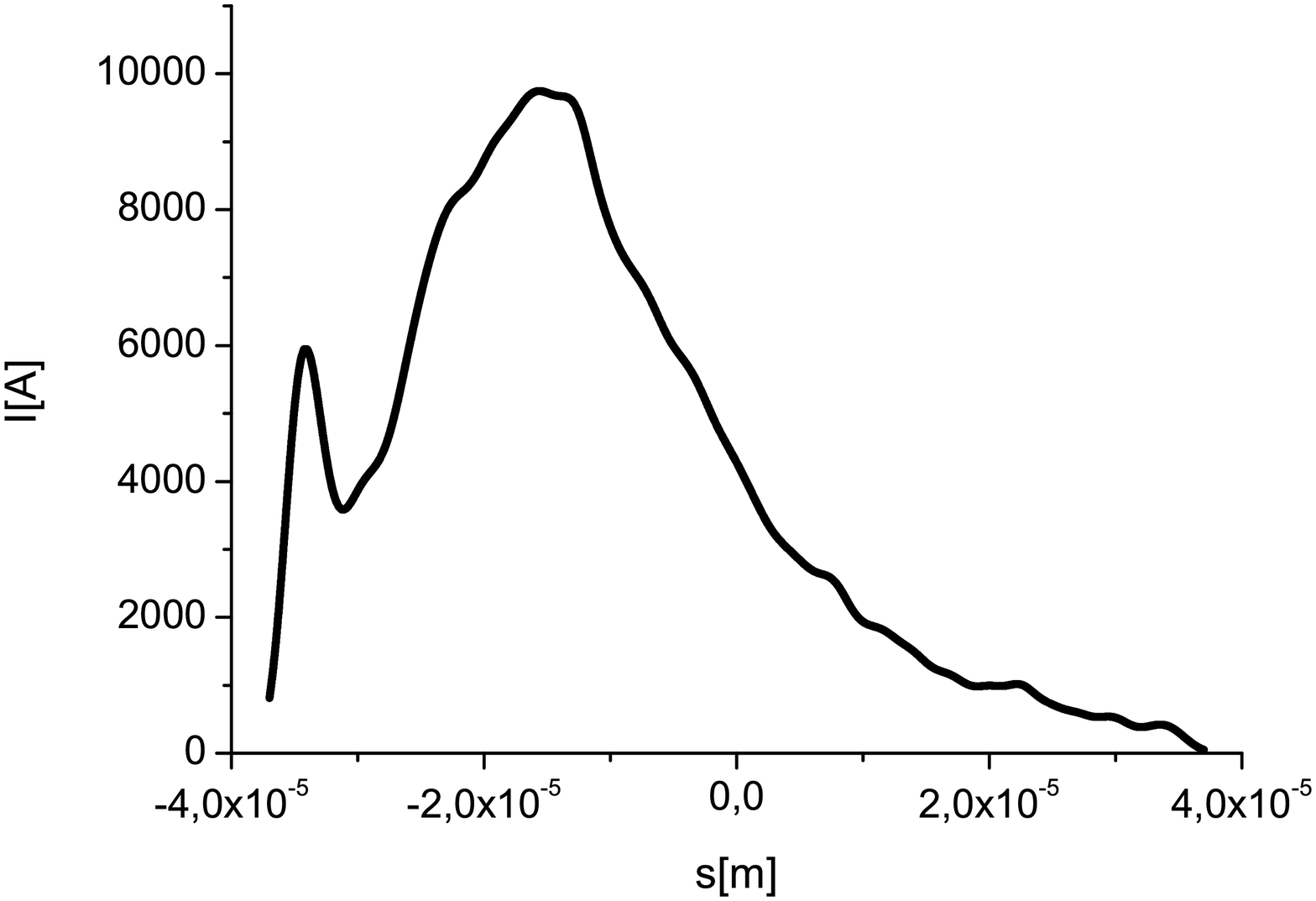}
\includegraphics[width=0.5\textwidth]{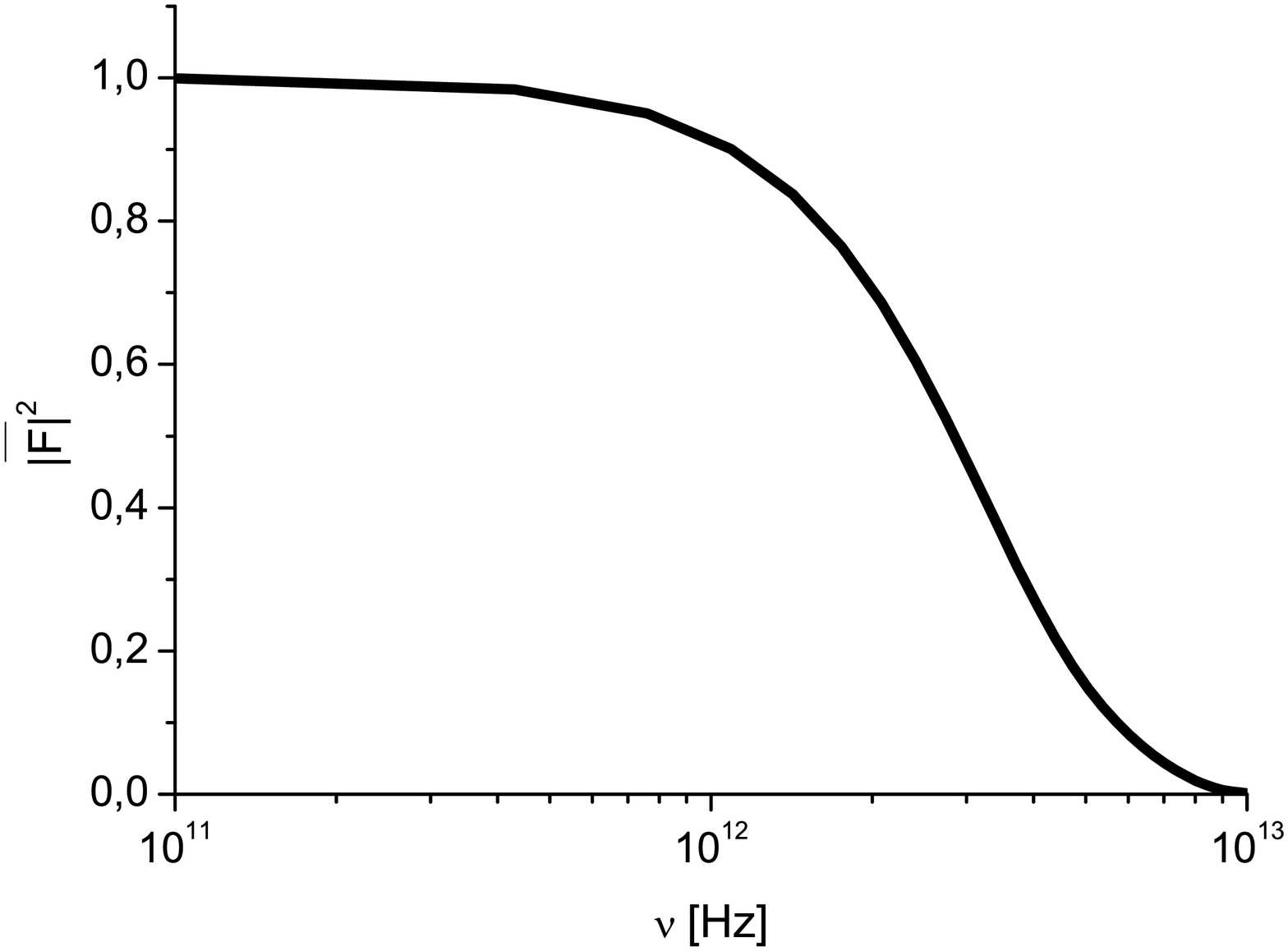}
\caption{Left: Electron beam current profile at the European XFEL,
optimized for THz pump/X-ray probe experiments. Right: Squared
modulus of the electron beam form factor, $|\bar{F}|^2$,
corresponding to the current profile on the left plot.}
\label{curform}
\end{figure}
Let us consider the practically important case of an electron bunch
with non-Gaussian shape similar to that, which can be used to drive
the European XFEL. Fig. \ref{curform} (left) shows the current
distribution along the bunch. The nominal charge is $1$ nC. The
electron bunch has a complicated shape, which is reflected in the
square modulus of the form factor shown in Fig. \ref{curform}
(right). In the case of the European XFEL, the squared of the bunch
form factor modulus falls off rapidly for wavelengths shorter than
$0.06$ mm. At the opposite extreme, the dependence of the form
factor on the exact shape of the electron bunch is rather weak and
can be ignored for wavelengths longer than $0.1$ mm.

We assume an electron beam moving along z-axis inside the
axisymmetric iris line through the screens, Fig. \ref{T2}. As
already remarked, in reference \cite{OURT} we developed a theory of
such kind of THz source using Veinstein impedance boundary
conditions and Green's function approach to solve the field
equations. We obtained the following expressions for the field at
the mirror position:

\begin{eqnarray}
\widetilde{E}_x(r,\phi) && = -\frac{2 i N_e e \bar{F}(\omega)
(1-\Delta) (1-2\Delta) L \cos(\phi) }{\omega} \sum_{k=1}^{\infty}
\frac{\nu_{1k}J_1(\nu_{1k}(1-\Delta)r/a)}{a^3 J_0^2(\nu_{1k})} \cr
&& \times
{\mathrm{sinc}\left[\left(\frac{\omega}{2c\gamma^2}+\frac{c
\nu_{1k}^2(1-2\Delta)}{2\omega
a^2}\right)\frac{L}{2}\right]}\exp\left(-\frac{i c
\nu_{1k}^2(1-2\Delta)}{2\omega a^2}\frac{L}{2}\right)\label{etildex}
\end{eqnarray}
for the horizontal component and

\begin{eqnarray}
\widetilde{E}_y(r,\phi) && = -\frac{2 i N_e e \bar{F}(\omega)
(1-\Delta) (1-2\Delta) L \sin(\phi) }{\omega} \sum_{k=1}^{\infty}
\frac{\nu_{1k}J_1(\nu_{1k}(1-\Delta)r/a)}{a^3 J_0^2(\nu_{1k})} \cr
&& \times
{\mathrm{sinc}\left[\left(\frac{\omega}{2c\gamma^2}+\frac{c
\nu_{1k}^2(1-2\Delta)}{2\omega
a^2}\right)\frac{L}{2}\right]}\exp\left(-\frac{i c
\nu_{1k}^2(1-2\Delta)}{2\omega a^2}\frac{L}{2}\right)\label{etildey}
\end{eqnarray}
for the vertical one.

Here $\Delta = (1+i) \beta_0 \sqrt{c b/(4 \omega a^2)}$, $\gamma$ is
the relativistic factor, $L$ is the distance between upstream edge
screen and mirror. Coherence is accounted through the form factor
$\bar{F}(\omega)$.

Only non-azimuthal  symmetric modes turn out to be driven by the
uniform motion of the space-charge distribution in the axisymmetric
iris guide. From a physical view point this is a sound result. In
fact, radiation is related with the energy change of particles,
which takes place through the scalar product of the electric field
and velocity of the particles. Since the transverse velocity is
equal to zero in the edge radiation case, symmetric modes cannot
lead to any energy change of the electrons moving along the axis of
the axisymmetric iris guide. Let us focus on the fundamental non
symmetric mode only. Neglecting losses, for the moment, the
amplitude for the orthogonal polarization components of the field in
a Cartesian coordinate system is given by

\begin{eqnarray}
&&E_x = A_1 \cos(\phi) J_1(3.83 r/a) \cr && E_y = A_1 \sin(\phi)
J_1(3.83 r/a)~, \label{modes}
\end{eqnarray}
where $A_1$ is a constant for a certain longitudinal position, $r =
\sqrt{x^2 + y^2}$, and $\phi = \arctan(y/x)$.   It follows from
these equations that the direction of the electric field is radial
i.e. varies as a function of the transverse position. Therefore, for
the edge radiation case we have two separate amplitudes for two
orthogonal polarization directions. They are not azimuthal symmetric
because they depend, respectively, on $x/r$ and $y/r$, that is on
the cosine and on the sine of the azimuthal angle. Only if one sums
up the intensity patterns referring to the two polarization
components one obtains the azimuthal symmetric intensity
distribution

\begin{figure}[tb]
\includegraphics[width=1.0\textwidth]{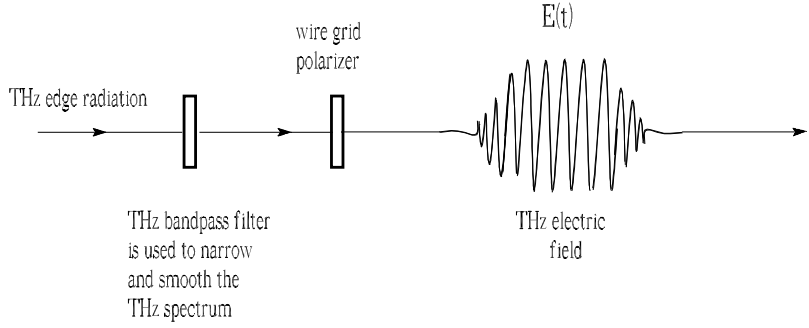}
\caption{Monochromatization and polarization filtering by wire grid
polarizer for producing linearly polarized THz radiation.}
\label{T3}
\end{figure}
\begin{eqnarray}
I \sim A_1^2 J_1^2( 3.83 r/a)~. \label{inten}
\end{eqnarray}
Edge radiation is characterized by broad spectrum and radial
polarization. Many practical applications require control of the
bandwidth and of the polarization of the THz pulse. The formation of
the THz edge radiation pulse usually involves monochromatization and
polarization filtering by wire-grid polarizer for producing linearly
polarized THz radiation, Fig. \ref{T3}.

Eq. (\ref{etildex}) and Eq. (\ref{etildey}) can be used as a basis
to calculate the energy per unit spectral interval per unit surface

\begin{eqnarray}
\frac{d W}{d\omega dS}&& = \frac{c}{4\pi^2}
\left(\left|\widetilde{E}_x\right|^2 +
\left|\widetilde{E}_y\right|^2\right)~.\label{eneee}
\end{eqnarray}
In our case of interest, the distance between sample and extracting
mirror, $L_s$, is much longer compared with the length of the edge
radiation setup. Integrating over transverse coordinates over the
iris hole, and assuming $L_s \gg L$ the energy of the radiation
pulse at the sample position is given by \cite{OURT}

\begin{eqnarray}
W && =   \frac{ e^2 L^2 c }{\pi \omega a^4} N_\mathrm{e}^2
\left|\bar{F}(\omega)\right|^2 \sum_{k=1}^{\infty}
\frac{\nu_{1k}^2}{J_0^2(\nu_{1k})} \exp\left(-\frac{
\nu_{1k}^2\beta_0 c^{3/2} b^{1/2} }{ \omega^{3/2} a^3}L_s\right) \cr
&& \times \mathrm{sinc}^2\left[\frac{L}{4}\left(\frac{ \omega}{c
\gamma^2}+\frac{ \nu_{1k}^2 c}{ \omega {a}^2}\right)\right]
\frac{\Delta \omega}{\omega}~.\label{energy}
\end{eqnarray}
Here we assumed that the bandwidth of spectral filter $\Delta
\omega/\omega \ll 1$ is small enough to neglect the dependence on
$\omega$ in Eq. (\ref{etildex}) and Eq. (\ref{etildey}).

\section{\label{tre} Scheme for generating THz edge radiation at the European
XFEL}

\begin{figure}[tb]
\includegraphics[width=1.0\textwidth]{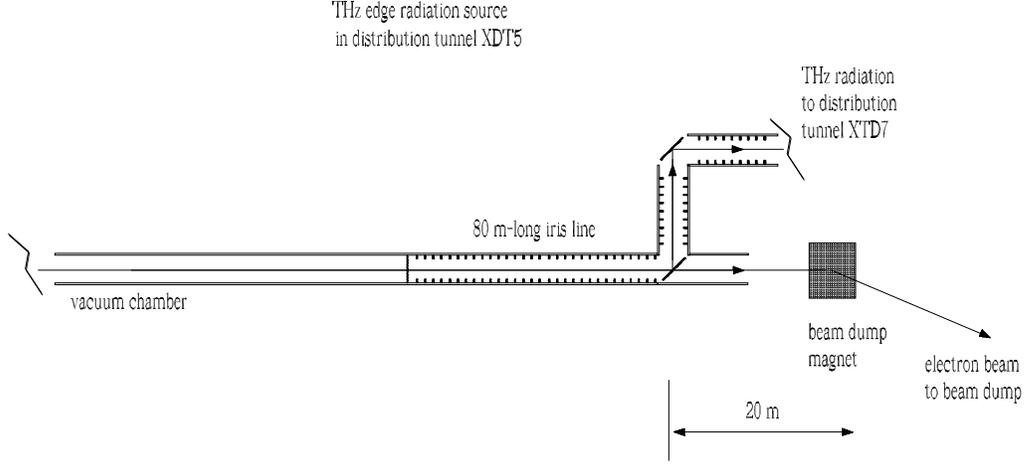}
\caption{Installation of the setup at the European XFEL.} \label{T9}
\end{figure}
\begin{figure}[tb]
\includegraphics[width=1.0\textwidth]{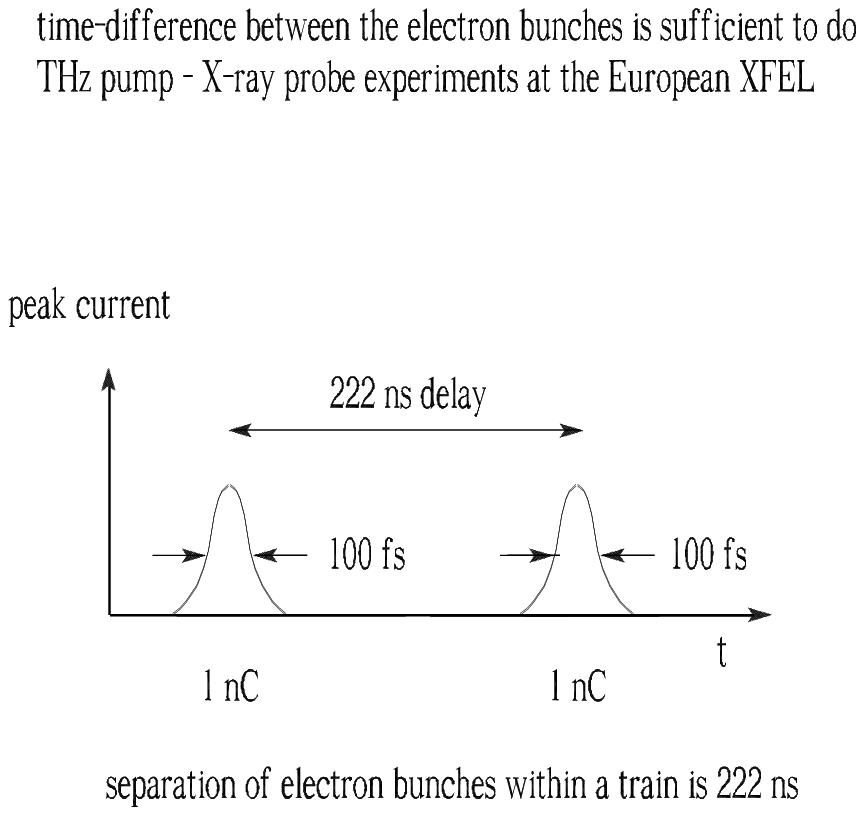}
\caption{Bunch structure in a macropulse for the baseline
multi-bunch operation mode at the European XFEL.} \label{T10}
\end{figure}
\begin{figure}[tb]
\includegraphics[width=1.0\textwidth]{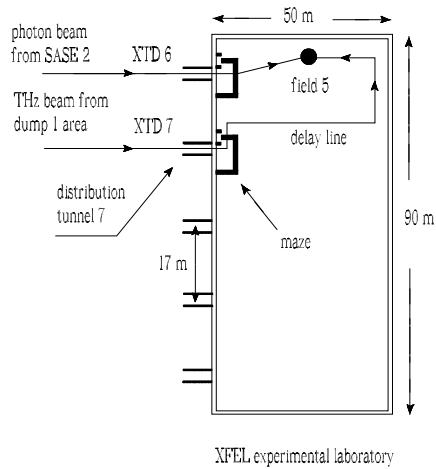}
\caption{Sketch of a possible arrangement for a THz delay line in
the experimental hall.} \label{T11}
\end{figure}
\begin{figure}[tb]
\includegraphics[width=1.0\textwidth]{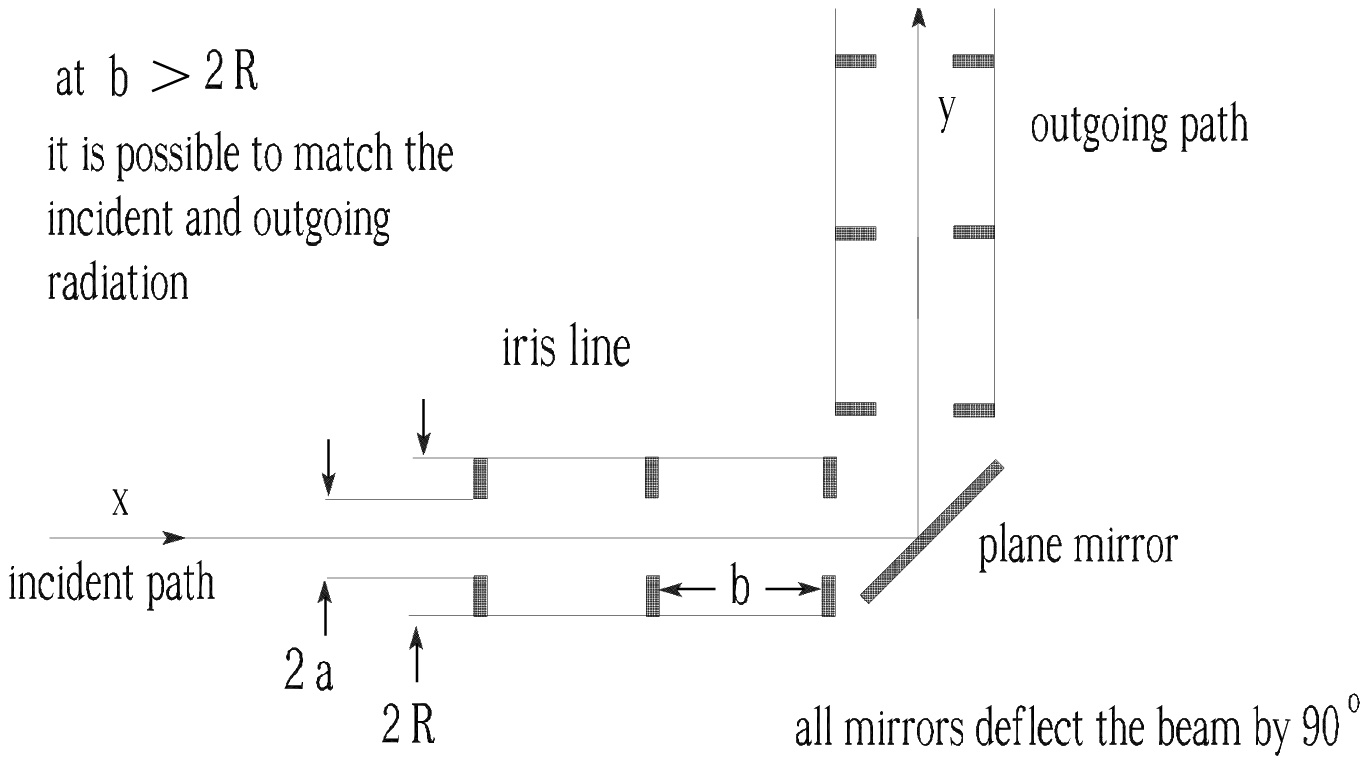}
\caption{Geometry of a transmission-line turn.} \label{T7}
\end{figure}
The THz edge radiation source proposed in this paper is compatible
with the layout of the European XFEL and can be realized with little
efforts and minimal modifications to the baseline setup. The vacuum
chamber equipped with iris line and outcoupling system can be
installed in the unoccupied straight vacuum line upstream of the
first electron beam dump, Fig. \ref{T9}.

The transport of the THz beam to the experimental hall can be
performed exploiting the use of the tunnel XTD7, as reproduced
schematically in Fig. \ref{T9}. The THz transmission line
transporting the THz radiation introduces a path-delay of about 20 m
with respect to the path of X-ray pulse from the SASE2 undulator.
Since THz pump/X-ray probe experiments should be enabled, we propose
to exploit the natural bunch spacing within a train, Fig. \ref{T10}.
The THz and X-ray pulses should be synchronized by a THz delay line
installed in experimental hall, Fig. \ref{T11}.

The THz transmission line includes at least ten $90$-degrees turns,
and will exploit plane mirrors as functional components. If the pipe
of the transmission line has a diameter smaller than  the distance
between the irises it is possible to match incident and outgoing
radiation without extra losses in these irregularities, Fig.
\ref{T7}.

\begin{figure}[tb]
\includegraphics[width=1.0\textwidth]{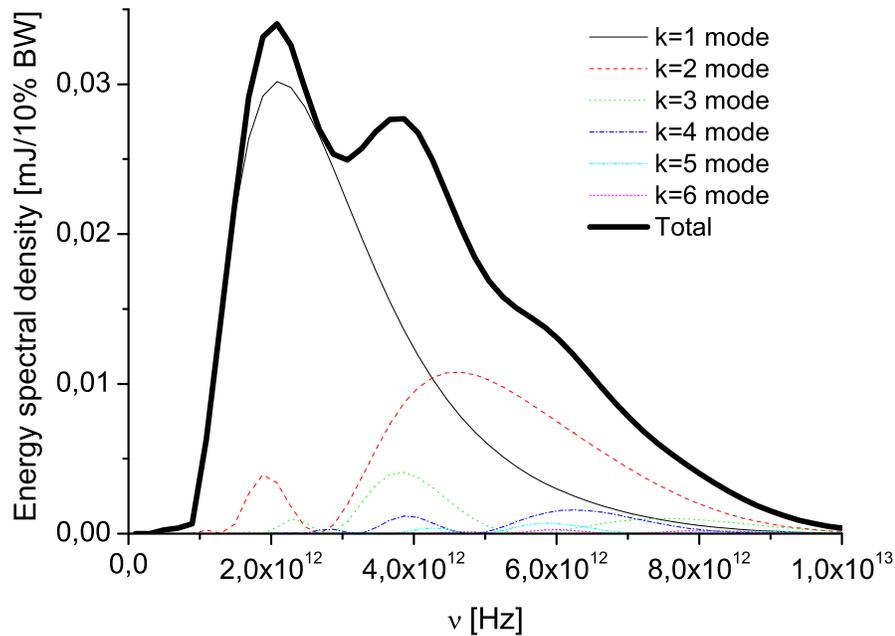}
\caption{Edge radiation pulse energy spectral density as a function
of frequency transported at the sample position. Partial
contributions of individual modes of the circular iris guide are
illustrated. The sample is set at $L_s = 250$ meters away from the
extracting mirror. The curves are calculated with analytical
formulas in Eq. (\ref{etildex}), Eq. (\ref{etildey}) and Eq.
(\ref{eneee}), including losses as in Eq. (\ref{loss}). Here $N_e =
6.4 \cdot 10^9$ ($1$ nC), $L = 80$ m, $b = 30$ cm, $2a = 15$ cm. The
bunch form factor used is shown on the right plot of Fig.
\ref{curform}.} \label{enepulse}
\end{figure}
It is possible to calculate the energy spectral density at the
sample position. The bunch form factor considered here is given in
Fig. \ref{curform}. We set a total length of the transport iris line
of $250$ m. The energy losses in the line are accounted for. The
energy spectral density at the exit of transport iris line as a
function of the frequency $\nu = \omega/(2 \pi)$ is shown in Fig.
\ref{enepulse}. The curve is calculated with analytical formulas in
Eq. (\ref{etildex}), Eq. (\ref{etildey}) and Eq. (\ref{eneee})
including losses as in Eq. (\ref{loss}). The first four modes and
the total are shown. Here $N_e = 6.4 \cdot 10^9$, corresponding to 1
nC charge, $b = 30$ cm, $2 a = 15$ cm, $L = 80$ m, and $L_s = 250$
m.

The maximum value of the energy spectral density is achieved at
$\lambda \simeq 150 ~\mu$m. When the bandpass filter is tuned to
this value of $\lambda$, the expression for the total edge radiation
pulse energy at the sample can be written in the form

\begin{eqnarray}
W[\mathrm{mJ}] \simeq 0.35 \frac{\Delta \omega}{\omega}~. \label{WW}
\end{eqnarray}
The energy spectral density as a function of frequency exhibits a
low frequency cutoff due to losses in the transport line and a high
frequency cutoff due to electron bunch form factor suppression.

\begin{figure}[tb]
\includegraphics[width=1.0\textwidth]{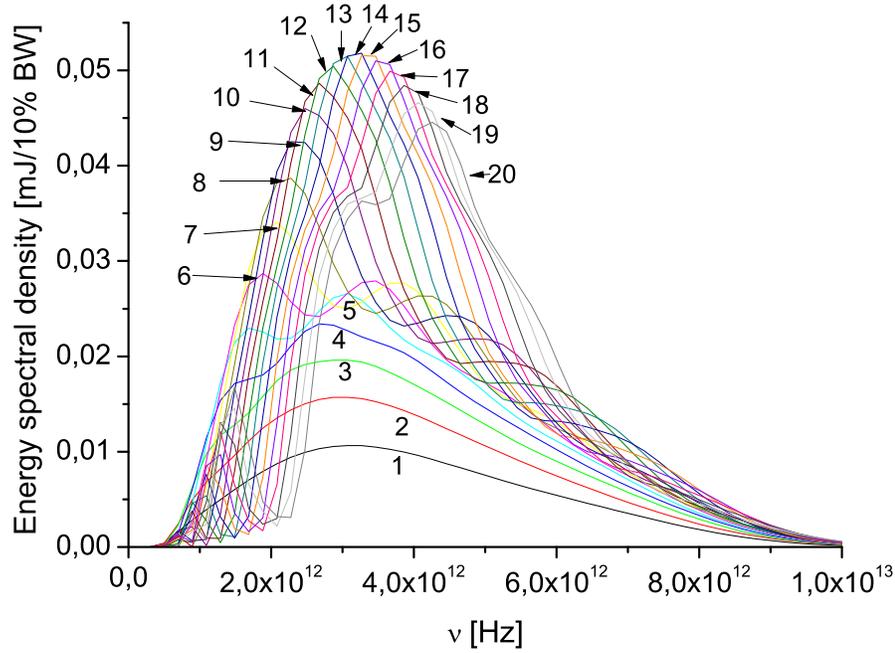}
\caption{Scan of the dependence on $L$ of the edge radiation pulse
energy spectral density transported at the sample position ($L_s =
250$ m downstream of the mirror) as a function of frequency. Here
$N_e = 6.4 \cdot 10^9$ ($1$ nC), $b = 30$ cm, $2a = 15$ cm. The
bunch form factor used is shown on the right plot of Fig.
\ref{curform}. The curves indicated with numbers from $1$ to $20$
refer to different values of $L$ starting from $L=20$ m and ending
with $L=210$ m, with a step of $10$ m.} \label{scanL}
\end{figure}
\begin{figure}[tb]
\includegraphics[width=1.0\textwidth]{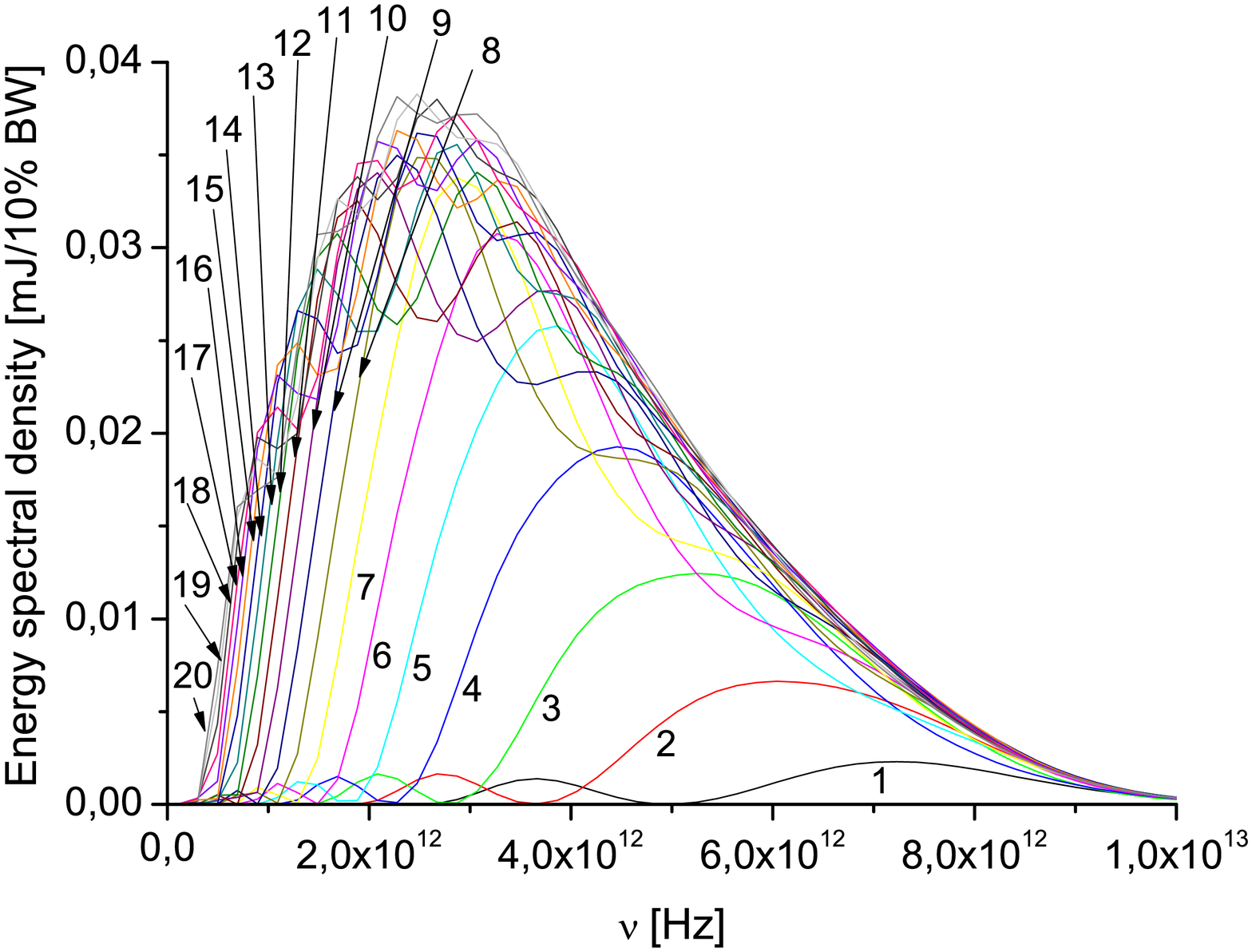}
\caption{Scan of the dependence on $a$ of the edge radiation pulse
energy spectral density transported at the sample position ($L_s =
250$ m downstream of the mirror) as a function of frequency. Here
$N_e = 6.4 \cdot 10^9$ ($1$ nC), $b = 30$ cm, $L = 80$ m. The bunch
form factor used is shown on the right plot of Fig. \ref{curform}.
The curves indicated with numbers from $1$ to $20$ refer to
different values of $a$ starting from $2 a=6$ cm and ending with $2
a=25$ cm, with a step of $1$ cm.} \label{scanA}
\end{figure}
We studied the influence of the edge radiation setup length $L$ and
of the iris hole radius $a$ on the operation of the THz source. Fig.
\ref{scanL} and Fig. \ref{scanA} show the dependence of the THz
pulse energy spectral density on the frequency for different values
of these two parameters.  The nontrivial behavior of the energy
spectral density reflects the complicated parametric dependence in
Eq. (\ref{energy}).

Finally, we note that according  to our calculations in Fig.
\ref{enepulse}, one obtains about $0.12$ mJ total THz radiation
pulse energy at the sample position at a wavelength around $0.075$
mm with a filter bandwidth $\Delta \omega/\omega  \sim 0.5$. For one
cycle in a pulse, corresponding to $50 \%$ spectral
bandwidth\footnote{Analysis of  Eq. (\ref{kzzz2}) shows that the
phase velocity of the radiation field of the iris guide mode is
larger than the velocity of light and depends on the frequency as
$\Delta k_z \sim 1/\omega$ for large values of the Fresnel number $N
\gg 1$. In other words, the THz radiation pulse propagates through
the transport line with a certain group velocity dispersion, and the
shape of the envelope broadens. In our case of interest, the
broadening is about $10 \%$, and can be neglected.}, one obtains
about 0.5 GW peak power level. By highly focusing this THz beam, one
will approach the high field limit of 1 V/atomic size.

\section{\label{quattro} European XFEL radiation parameters for THz pump/X-ray probe experiments}

The electron beam formation system of the European XFEL accelerator
complex operates with nominal charge of $1$ nC. The initially 2 mm
(rms) long bunch is compressed in three magnetic chicanes by a
factor of $4$, $8$ and $4$ respectively, to achieve a peak current
of $5$ kA.

The design of our THz source is also based on the exploitation of
electron bunches with nominal charge of $1$ nC. However, in order to
optimize the THz source performance, the electron bunch should be
made about two times shorter than in the nominal mode of operation.
In this case, the peak current increases by a factor two. Operation
beyond the design parameters leads to some degradation of the
electron beam quality. The operation of the proposed THz source is
insensitive to the emittance and energy spread of the electron
bunch. However, one should examine the performance of the electron
bunches in terms of X-ray SASE pulses generation, since THz
pump/X-ray probe experiments are based on such pulses.

\begin{figure}[tb]
\includegraphics[width=0.5\textwidth]{curr.eps}
\includegraphics[width=0.5\textwidth]{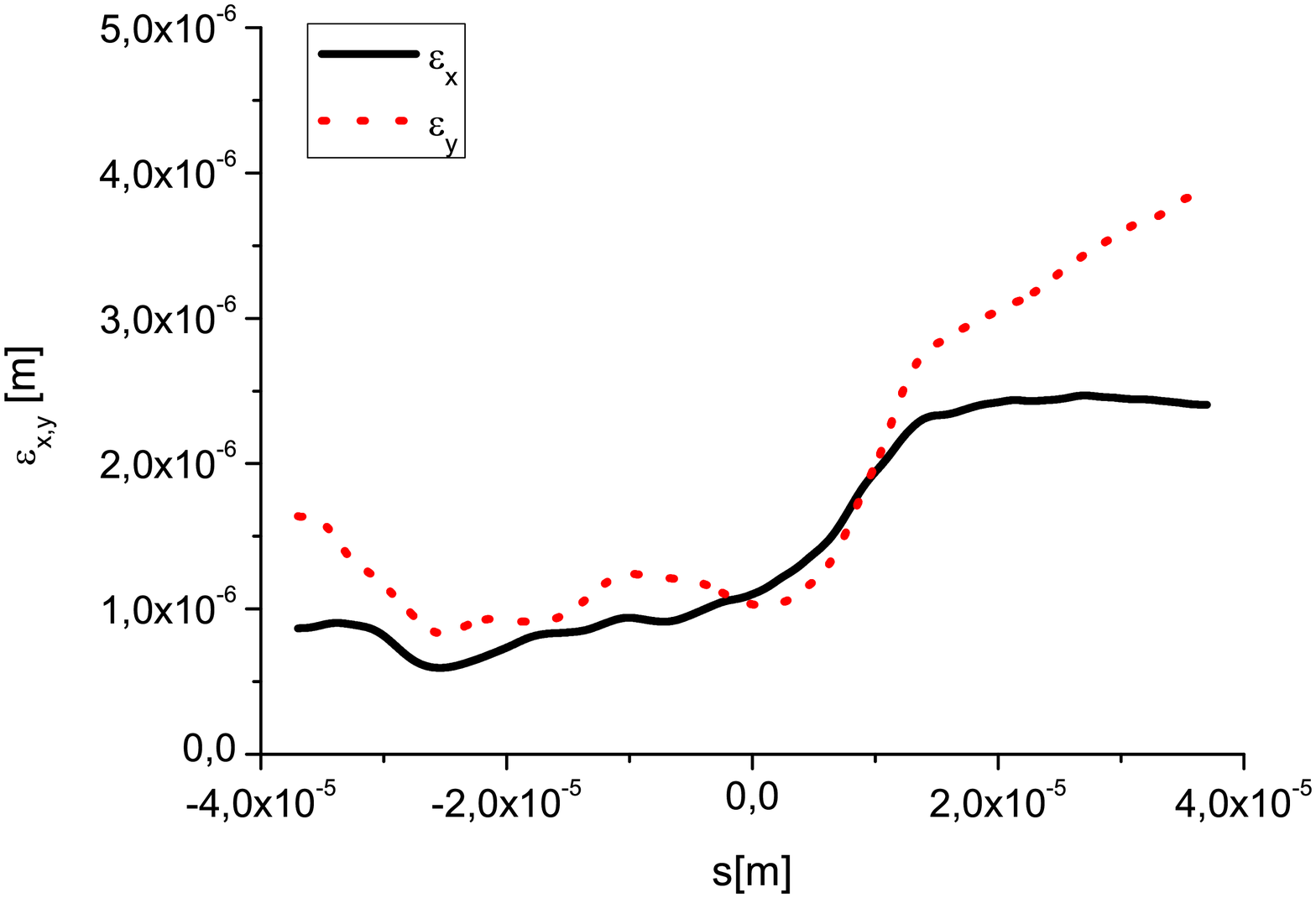}
\includegraphics[width=0.5\textwidth]{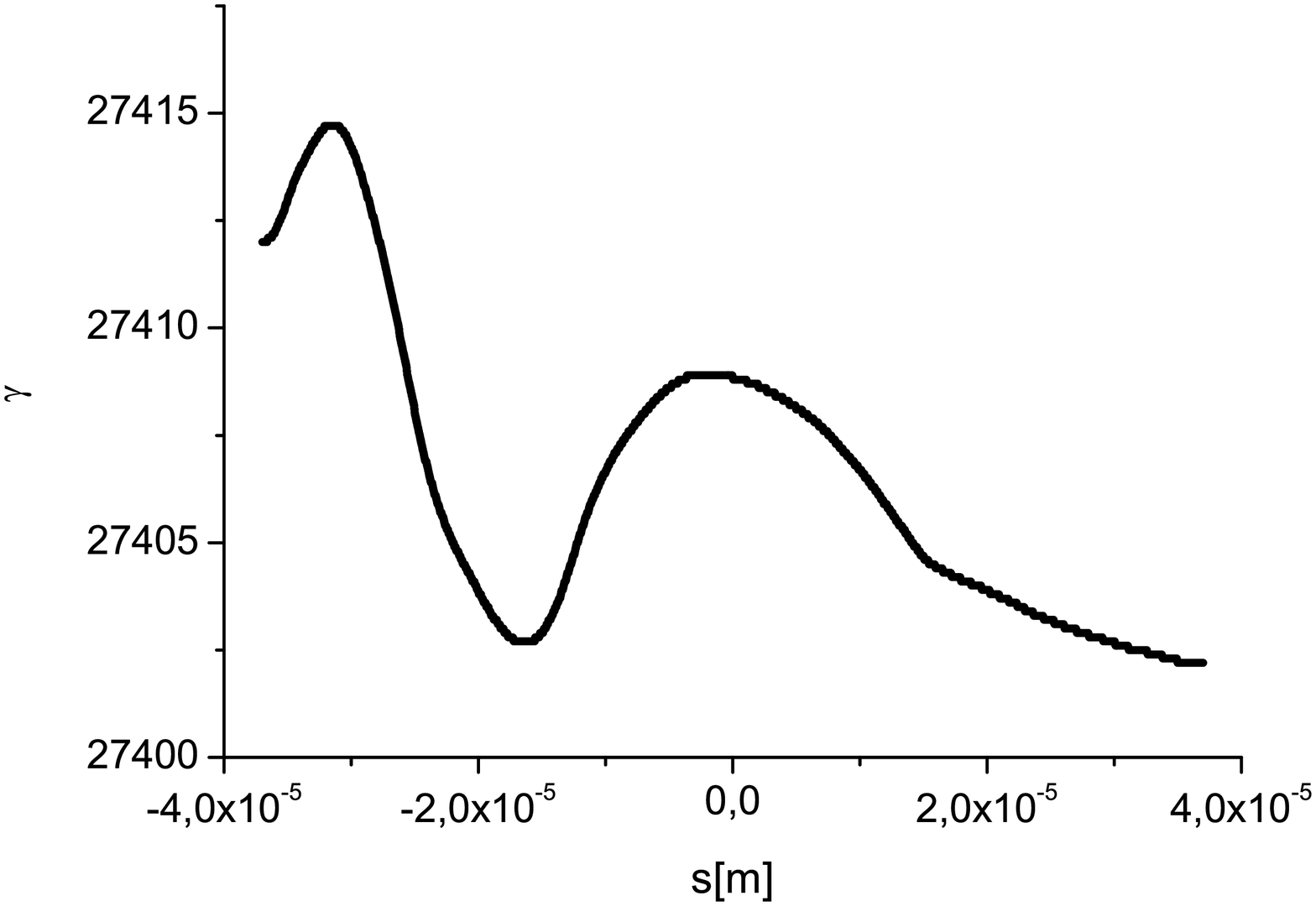}
\includegraphics[width=0.5\textwidth]{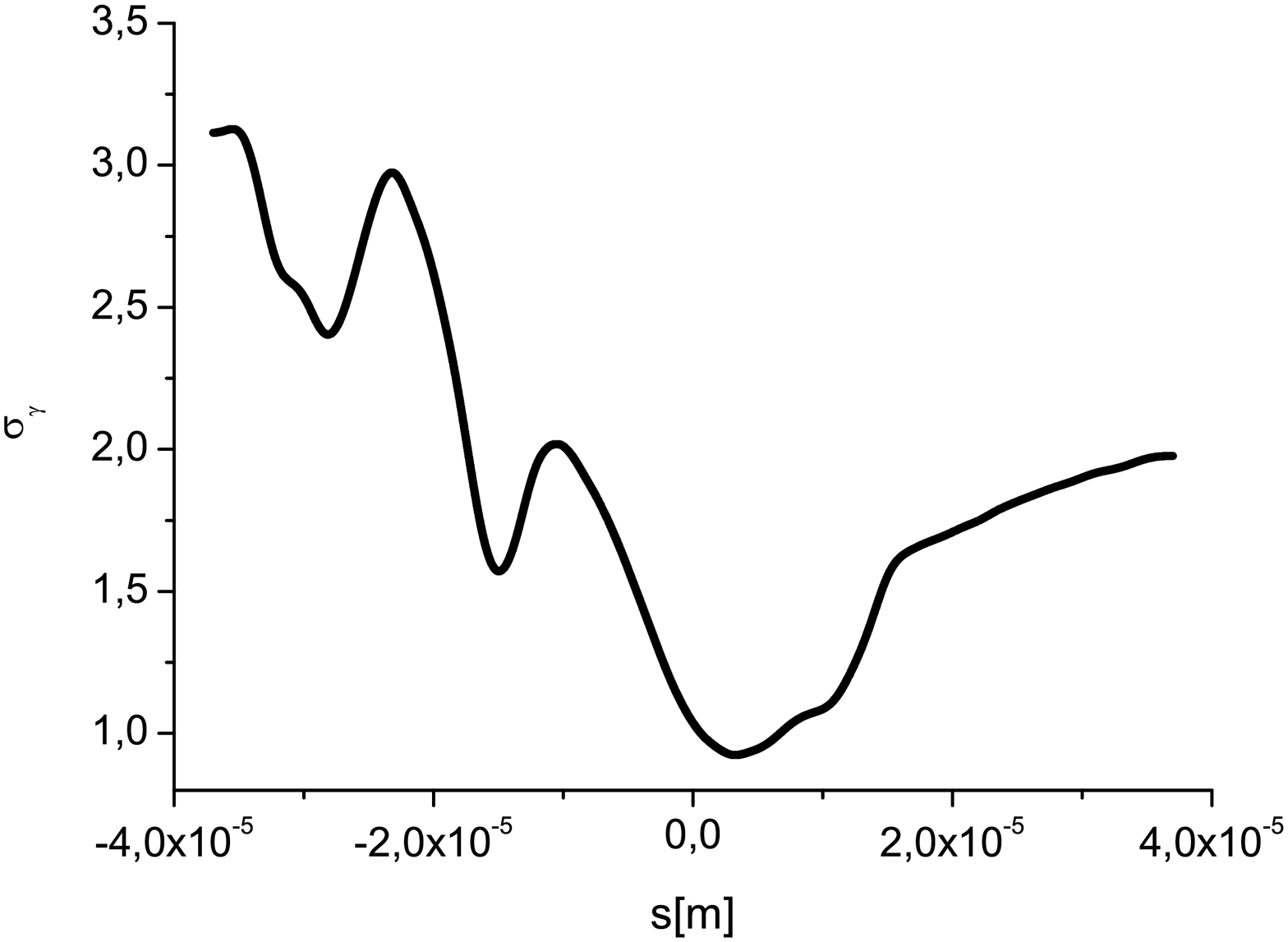}
\caption{Results from electron beam start-to-end simulations at the
entrance of SASE2. Top Left: Current profile. Top Right: Normalized
emittance as a function of the position inside the electron beam.
Bottom Left: Energy profile along the beam. Bottom right: Electron
beam energy spread profile.} \label{SASEs2E}
\end{figure}
First, we ran start-to-end simulations for the beam formation system
\cite{ZAGO}, yielding the beam characteristics summarized in Fig.
\ref{SASEs2E} in terms of current, emittance, energy and energy
spread distribution along the electron bunch at the entrance of the
SASE2 undulator. Subsequently, we studied the performance of the
electron bunch with the help of the FEL code GENESIS 1.3 \cite{GENE}
running on a parallel machine. Here we will present results based on
a statistical analysis consisting of $100$ runs.

\begin{figure}[tb]
\includegraphics[width=1.0\textwidth]{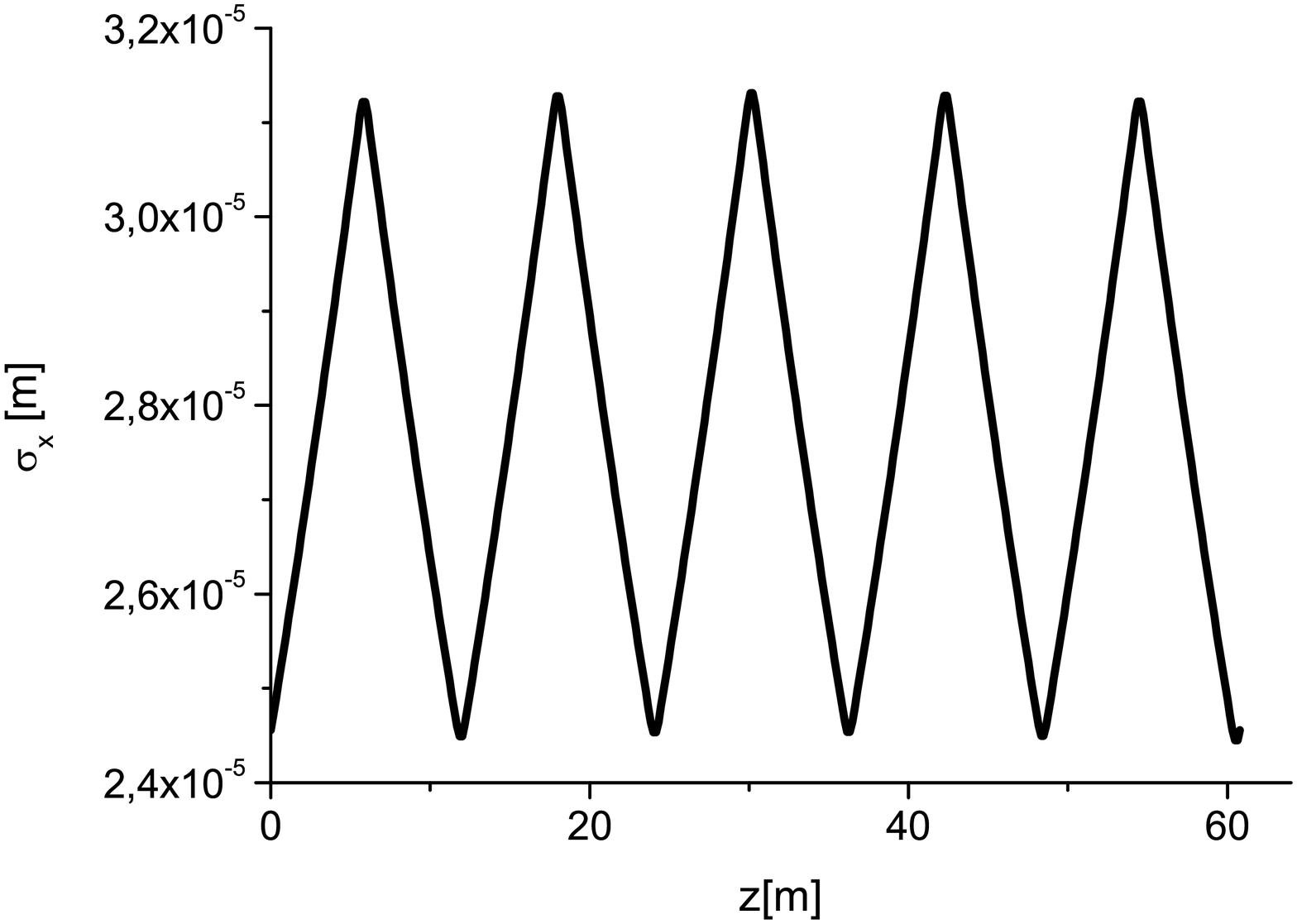}
\caption{Evolution of the rms horizontal beam size as a function of
the distance along the SASE undulator calculated through GENESIS.
These plots refer to the position within the bunch where the current
is maximal. } \label{sigperp}
\end{figure}
Fig. \ref{sigperp} shows the evolution of the rms horizontal beam
size as a function of the distance inside the SASE2 undulator. The
figure shows the evolution for the position of
maximal current in the bunch. 

\begin{figure}[tb]
\includegraphics[width=0.5\textwidth]{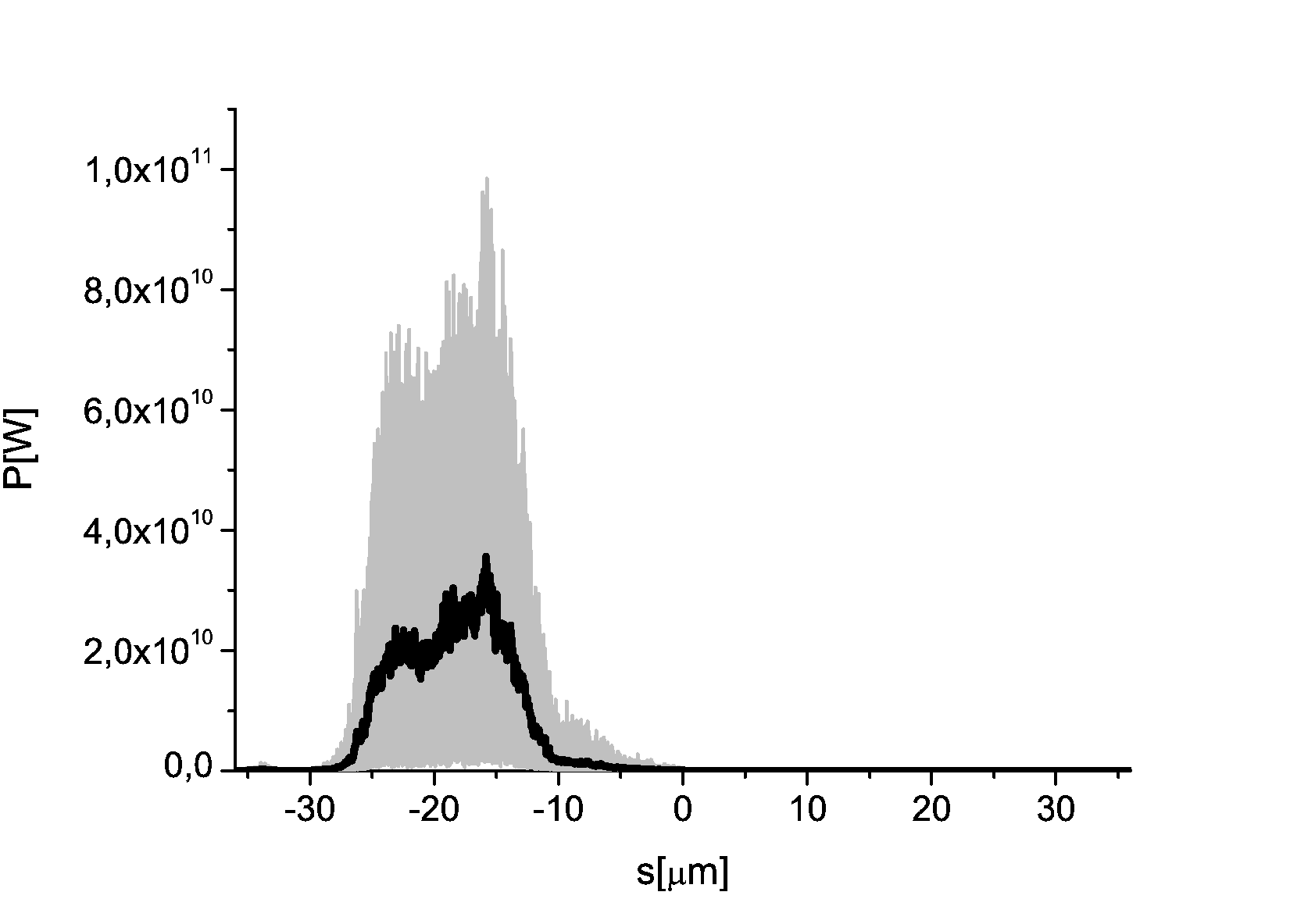}
\includegraphics[width=0.5\textwidth]{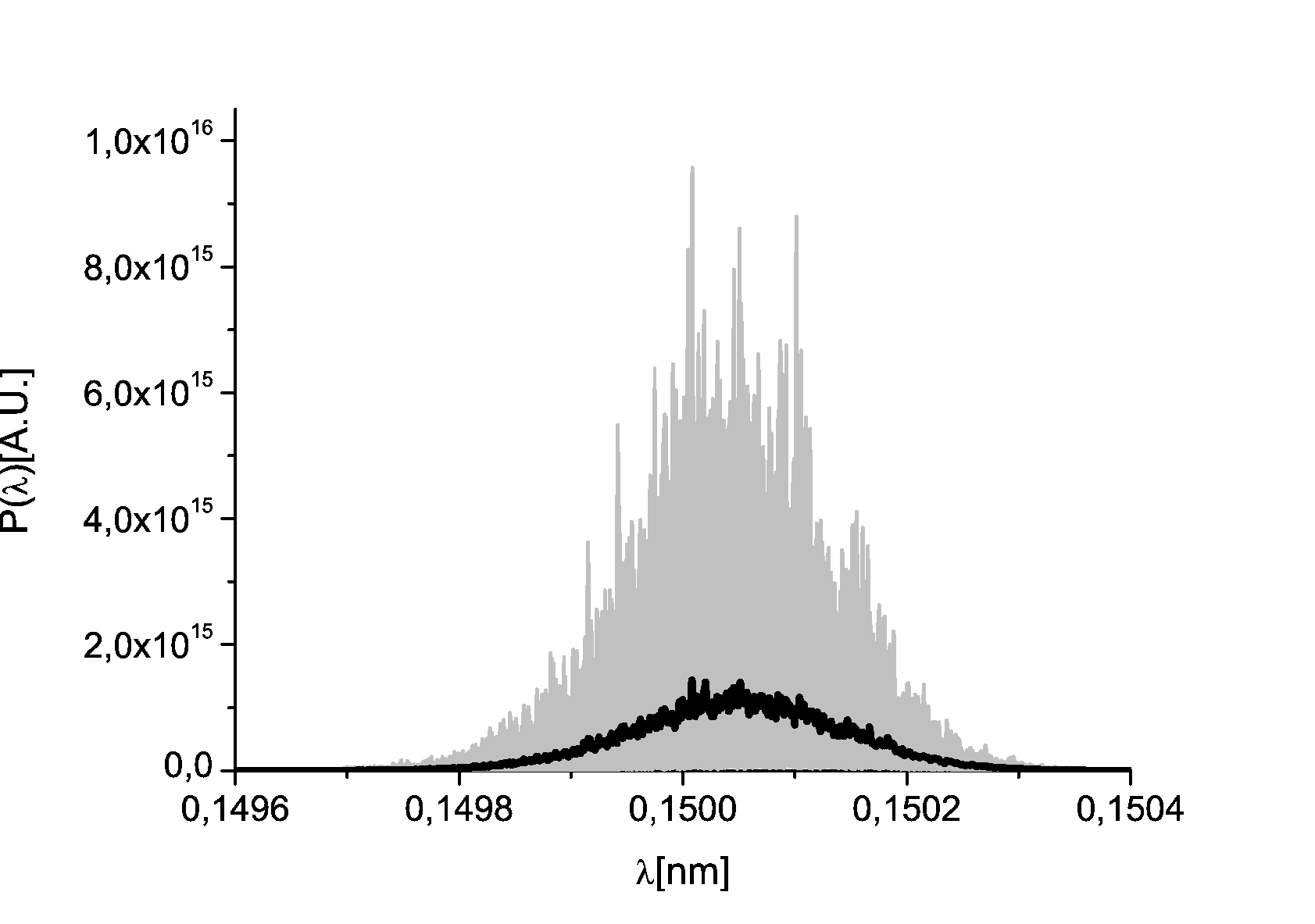}
\caption{Left plot: SASE power at the exit of the first $10$ FEL
undulator cells. Right plot: SASE spectrum at the exit of the first
$10$ FEL undulator cells. Grey lines refer to single shot
realizations, the black line refers to the average over a hundred
realizations.} \label{SASEout}
\end{figure}
The SASE2 undulator is composed of $35$ cells.  The SASE output
after the first $10$ undulator cells is shown in Fig. \ref{SASEout}.
On the left plot we show the output power, while on the right plot
one can see the spectrum around the the fundamental at $1.5$
Angstrom. Grey lines refer to single shot realizations, the black
line refers to the average over a hundred realizations. The output
characteristics are in line with the expected baseline performance,
demonstrating an average power of several tens of GW, and a relative
bandwidth in the $10^{-3}$ range.

\begin{figure}[tb]
\includegraphics[width=0.5\textwidth]{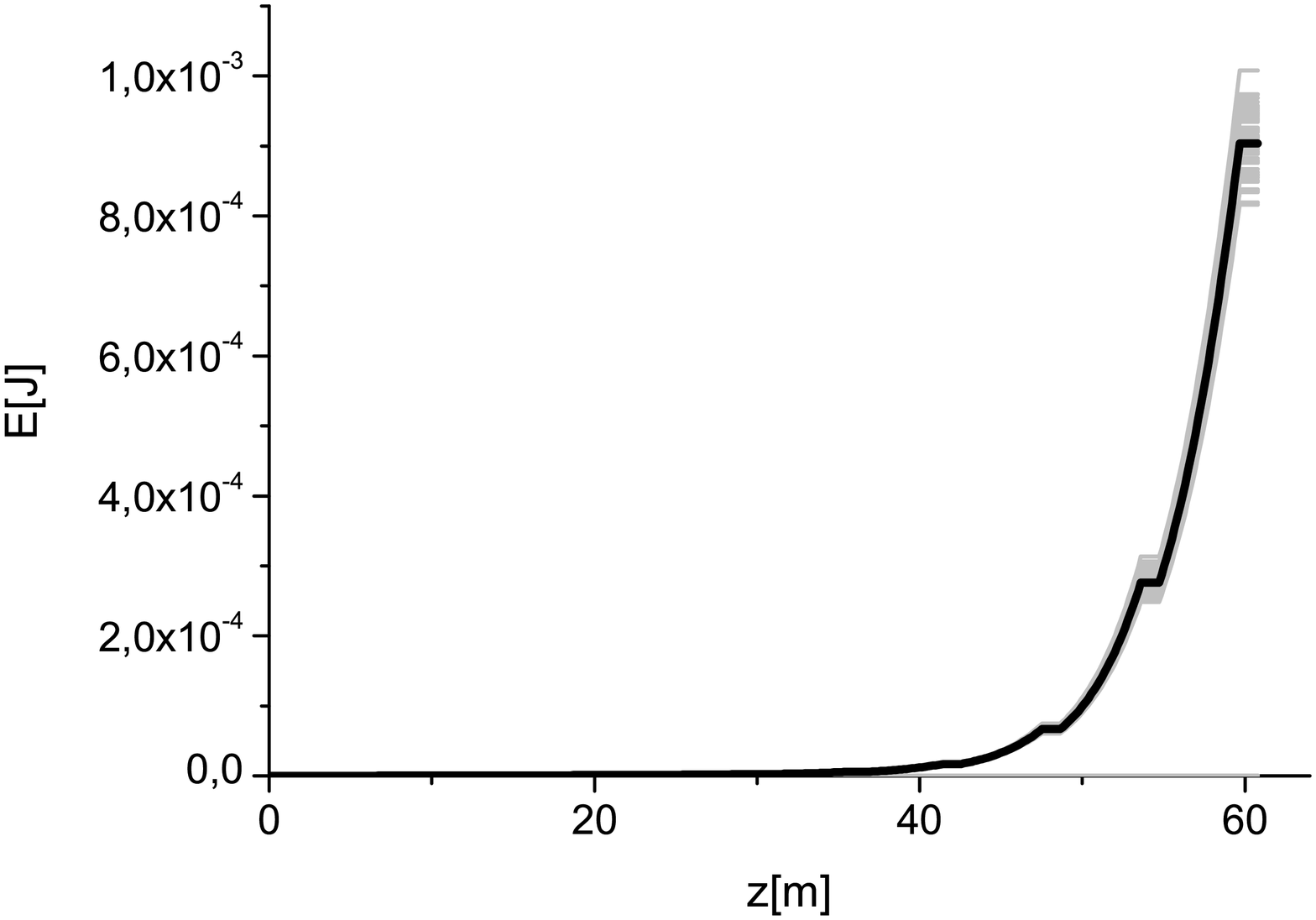}
\includegraphics[width=0.5\textwidth]{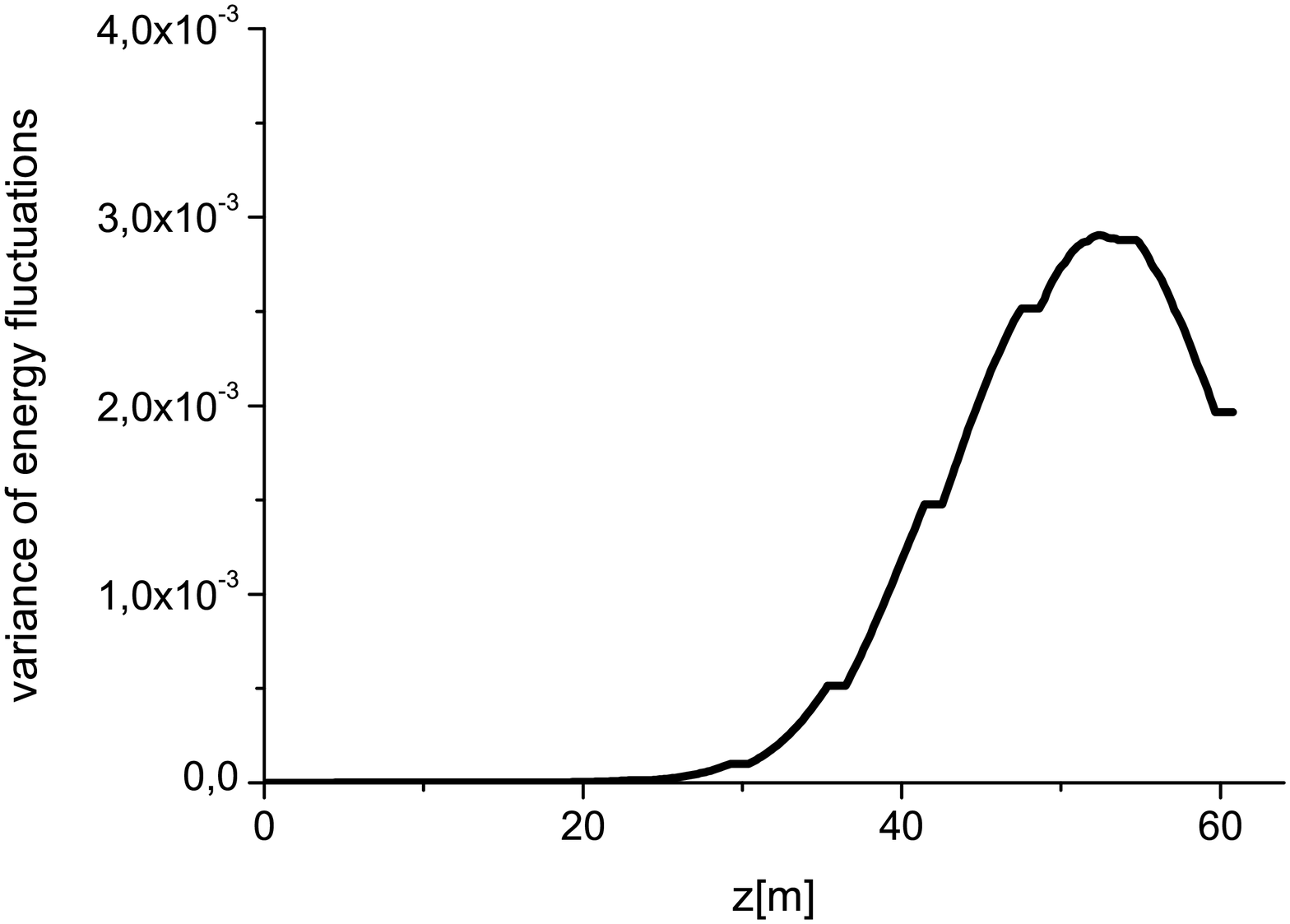}
\caption{Left plot: Evolution of the energy in the FEL pulse as a
function of the position in the undulator. Grey lines refer to
single shot realizations, the black line refers to the average over
a hundred realizations. Right plot: Variance of the energy
fluctuations.} \label{SASEevo}
\end{figure}

\begin{figure}[tb]
\includegraphics[width=0.5\textwidth]{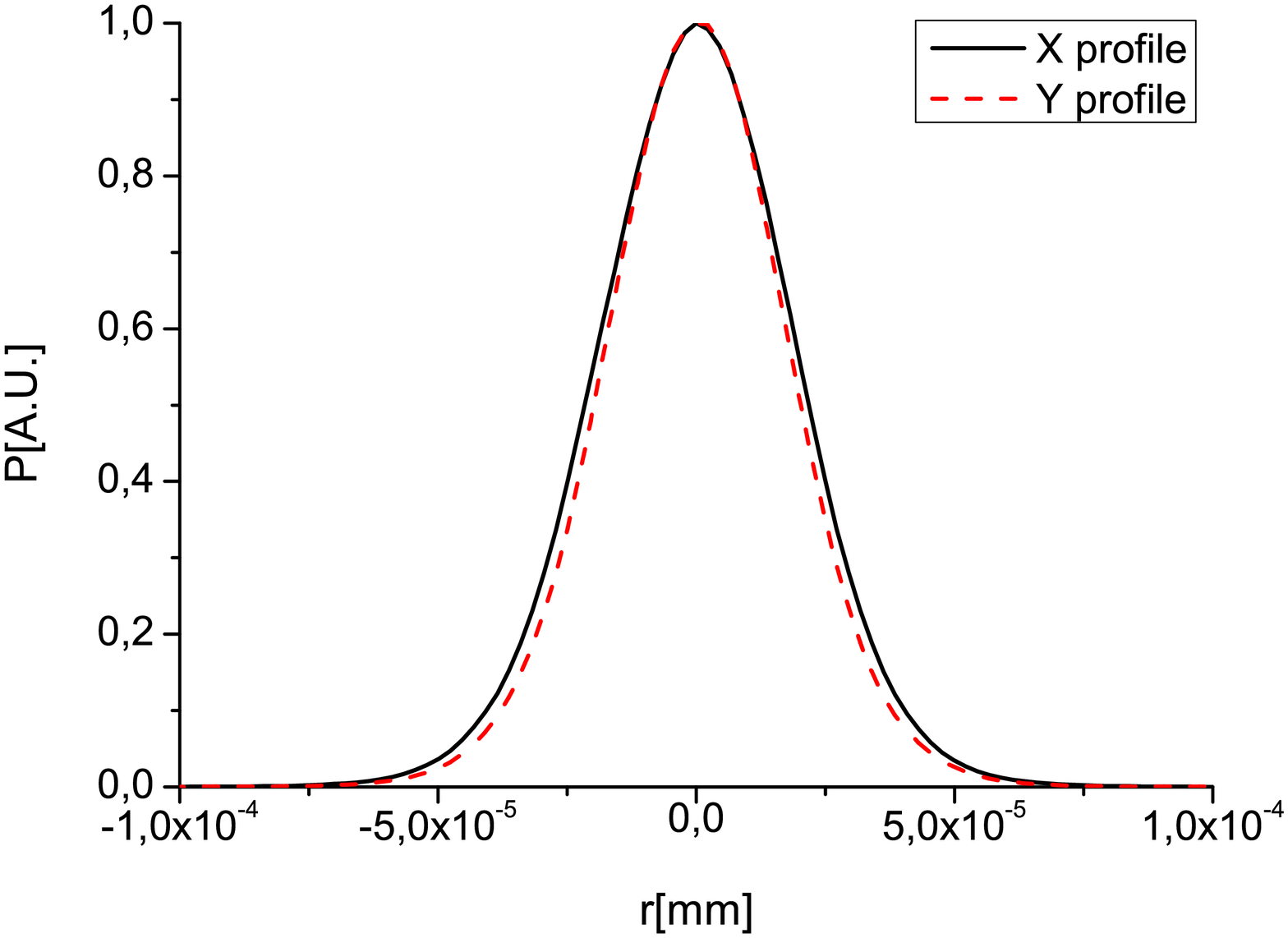}
\includegraphics[width=0.5\textwidth]{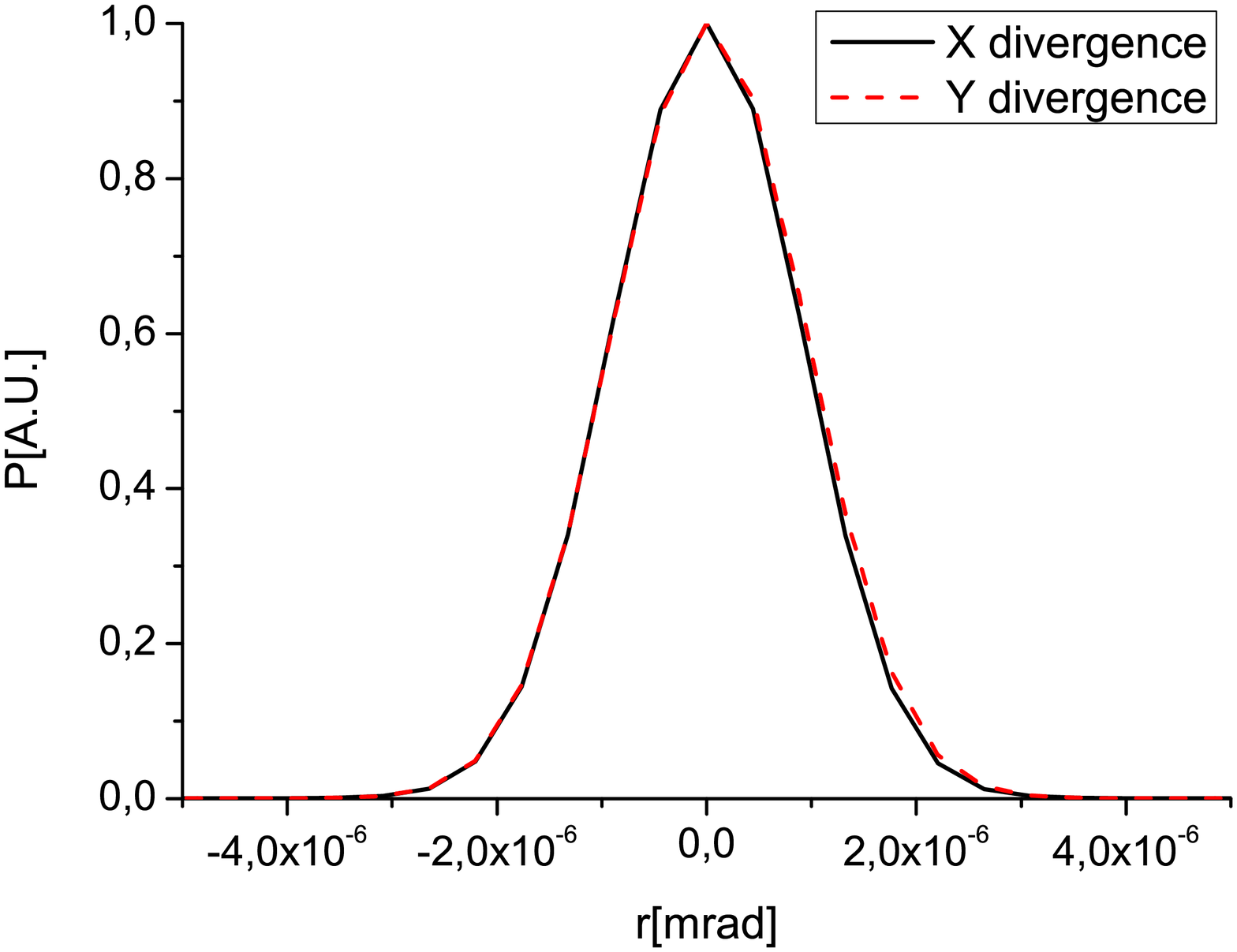}
\caption{Left plot: Transverse radiation distribution at the exit of
the FEL undulator. Right plot: Directivity diagram of the radiation
distribution at the exit of the FEL undulator.} \label{SASEspot}
\end{figure}
The left plot in Fig. \ref{SASEevo} shows the evolution of the pulse
energy as a function of the position inside the SASE2 undulator,
while the right plot shows the variance of the energy fluctuations.
As usual, fluctuations increase in the linear regime, and drop at
saturation. Finally, in Fig. \ref{SASEspot} we present the
transverse profile of the radiation distribution, obtained by
averaging over the statistical ensemble and the directivity diagram
of the radiation, obtained by Fourier transformation. Fig.
\ref{SASEspot} demonstrates a spot size of several tens of microns,
and a divergence of a few microradians.

Simulations presented in this Section show that the operation of
SASE2 with a $1$ nC bunch and a peak-current increase of a factor
two compared to the nominal mode of operation (also with a $1$ nC
bunch but with design parameters), leads practically to the same
peak power of the X-ray pulse that is in the order of $20$ GW. In
other words, the two times shorter bunch duration in our case study
has the simple consequence of a decrease in the X-ray pulse energy
of about a factor two, from around $2$ mJ in the design case to
around $1$ mJ in our case.

\section{\label{sei} Conclusions}

The accelerator complex at the European XFEL will produce ultrashort
electron bunches approaching sub-hundred fs duration. It is natural
to take advantage of these ultra-short bunches to provide coherent
THz radiation. THz radiation pulses can be generated by the spent
electron beam downstream of X-ray undulators. Any method relying on
the spent electron beam at the European XFEL should also provide, at
the same time, a way of transporting the radiation up to
experimental hall for a distance in the $300$-meters range.

Transmission of the THz beam can only be accomplished with quasi-
optical techniques. For the European XFEL case we propose to use an
open beam waveguide such as an iris guide. In order to efficiently
couple radiation into the iris transmission line, it is desirable to
match the spatial pattern of the source radiation to the propagating
mode of the transmission line. To solve the matching problem, we
propose to generate  the THz radiation directly within the iris
guide. The resulting electric field is found as superposition of the
iris waveguide modes, and is studied for the parameters case of the
European XFEL. We demonstrated that the maximally achievable field
strength at the sample is in the high field region of 1 V/atomic
size.

The present design describes a basic option for a THz source at the
European XFEL facility. A technical solution allowing for
significant performance enhancement of the output, of about an order
of magnitude, without significant changes of the baseline hardware,
seems also feasible. Such solution consists in the extension of the
proposed method to a higher charge mode of operation (up to $3$ nC),
which is within the possibilities of the European XFEL accelerator
complex. It is hoped that future research may extend the THz source
performance to a few GW power level.

\section{Acknowledgements}

We are grateful to Massimo Altarelli, Reinhard Brinkmann, Serguei
Molodtsov, Thomas Tschentscher and Edgar Weckert for their support
and their interest during the compilation of this work.

\end{document}